\documentclass[prb,twocolumn,superscriptaddress,showpacs]{revtex4}
\usepackage{amssymb}
\usepackage{txfonts}
\usepackage{bbm}
\usepackage{graphicx,graphics,color,epsfig}
\usepackage{appendix}
\usepackage{epstopdf}
\usepackage{appendix}

\begin{document}

\title{Competing ferromagnetic superconducting states in europium-based iron pnictides}

\author{Huai-Xiang Huang\footnote{hxhuang@shu.edu.cn}}
\affiliation{Department of Physics, Shanghai University, Shanghai 200444,
China}

\author{Yu-Qian Cao}
\affiliation{Department of Physics, Shanghai University, Shanghai 200444,
China}

\author{Xin Wan\footnote{xinwan@zju.edu.cn}}
\affiliation{Zhejiang Institute of Modern Physics, Zhejiang University, Hangzhou 310027, China}

\date{\today}
\begin{abstract}
In europium-based iron pnictides superconducting Fe-planes can be influenced by a Zeeman field originated from the neighboring Eu-planes.
The field tends to induce spin-density waves with a ferromagnetic average which coexists with the superconducting order by forming complementary patterns of the superconducting and magnetic order parameters in a Fulde-Ferrell-Larkin-Ovchinnikov phase and a two-dimensional textured-superconducting phase. The hard gap around the Fermi energy disappears in these fragile inhomogeneous superconducting states, which features, instead, V-shaped spin-resolved local density of states.
The inhomogeneous states are also competing with either a homogeneous superconducting or a homogeneous ferromagnetic state, manifesting the intertwining influences of the magnetic orders in Fe and Eu planes, the spin-density wave band structure, and the superconducting pairing order.

\end{abstract}

\maketitle

\section{Introduction}
In rare-earth compounds, such as REMo$_6$Se$_8$ (RE stands for rare-earth element), long-range ferromagnetic (FM) order tends to suppress superconducting (SC) order drastically~\cite{add1}. However, the coexistence of superconductivity (Fe-plane) and ferromagnetism (Eu-plane) has been observed in several EuFe$_2$As$_2$ systems~\cite{add2,7ja1,7ja2,cgh,vss,SZapf,kii,12}, in which superconductivity may be induced by mechanical pressure, isovalent substitution, and electron or hole doping.
Depending on pressure, doping, and temperature~\cite{add3,add4,add16,add17,add18}, the parent compound EuFe$_2$As$_2$ exhibits spin-density-wave (SDW) on Fe ions with different magnetic order on the Eu lattice.
Measurements of the neutron spin resonance mode and optical conductivity~\cite{kii} indicate that $s_\pm$-wave pairing symmetry is mostly favorable in Eu-based iron pnictides. The generic phase diagram of doped europium-based iron pnictides exhibits complex electronic phases, including a re-entrant spin glass phase~\cite{SZapf,add20} and a resistivity re-entrant~\cite{add21,add22,ja40,25,26,27} state associated with the long-range magnetic order of Eu$^{2+}$.

The interplay of the FM and SC order on the same plane may result in the Fulde-Ferrell-Larkin-Ovchinnikov (FFLO) state~\cite{f0}, which is a spatially varying SC state and can be stabilized by a Zeeman splitting due to either an external parallel magnetic field or an internal exchange field~\cite{f1,f2,f3,f33,f4,f5,f6,f7}. Although there is no undisputed experimental verification of its existence, iron-based superconductors have been suggested~\cite{cgh,andimp,Beyer} to support the FFLO state.

The newly synthesized RbEuFe$_4$As$_4$, which is hole doped as a whole, opens possibilities to tune the interaction between neighboring Eu layers.
Intriguingly, each Eu plane is ferromagnetically ordered below the SC critical temperature~\cite{cgh,vss}. To model the electronic properties on a Fe plane, we assume the coupling between the Eu and Fe planes is weak and replace the FM-ordered Eu planes by a Zeeman field $B$ without considering the feedback from the Fe planes. We also neglect the complex magnetic interaction of the stacked Eu planes along the $c$-axis. The compound has asymmetric arsenic ions above and below the Fe planes~\cite{kii}, so we adopt a realistic model containing two inequivalent Fe ions to investigate the effect of the FM order of the Eu planes on the superconductivity in the Fe planes.

With fixed average hole concentration, we calculate the zero-temperature phase diagram and demonstrate a subtle competition among the SC order, the magnetic response, and the kinetic energy.
As the Zeeman field is increased, the homogeneous SC (H-SC) phase gives way to the FFLO phase, then the so-called 2D textured-SC (TT-SC) phase~\cite{texture}, followed by the striped-FM phase, demonstrating the intricate interplay between superconductivity and magnetism. For a large range of parameters, inhomogeneous phases win energetically over homogeneous superconducting and/or ferromagnetic phases.
In addition to the patterns of the order parameters and the local density-of-states (LDOS), the competing phases can be distinguished by the electron occupation.

The rest of this paper is organized as follows. In Sec.~\ref{Sec2}, we introduce the theoretical model and the methods in our study. We show the results for order parameters and the zero-temperature phase diagram in Sec.~\ref{results}. We discuss the LDOS and the electron occupation in Sec~\ref{dos}, before we summarize in Sec.~\ref{final}.

\section{ Model and Method}\label{Sec2}

We start from a two-orbital four-band tight-banding Hamiltonian~\cite{m1,m2,s1,s11}
with an applied Zeeman field $B$:
\begin{eqnarray}
H_{t}=-\sum_{i\mu j\nu \sigma}t_{ i\mu, j\nu}  c^{\dag}_{i\mu\sigma}c_{j\nu\sigma}+{\rm h.\,c.}
- \sum_{i \mu \sigma}(\mu_0+\sigma B )c^{\dag}_{i\mu\sigma}c_{i\mu\sigma},
\end{eqnarray}
where $c^{\dag}_{i\mu\sigma}$ creates an electron with spin $\sigma$ in orbital $\mu$ at site $i$. $\mu_0$ is the chemical potential, from which all energies are measured. The average electron concentration is chosen to be $\langle n_{i} \rangle=1.8$ in all our calculations, such that the system is deep in the SC state in the absence of $B$.
We set the distance between nearest-neighbor (nn) Fe ions as the unit length and the nn hopping integral between the same orbitals to be $t_1 = 1$. The next nn (nnn) hopping integrals between the same orbitals mediated by the above and below As ions are $t_2$ and $t_3$, respectively, while the nnn hopping strength between different orbitals is $t_4$.
The hopping parameters are chosen to be $t_{2-4}=0.4,-2,0.04$ such that the resulting band structure is in good agreement with the ARPES and inelastic neutron scattering experiments~\cite{m1,m2}.

In the mean field framework, the SC Hamiltonian $H_{sc}$ and the interaction term $H_{int}$ can be written as~\cite{m2,s1,s11,s2,s3,s4}
\begin{eqnarray}
H_{sc}&=&\sum_{i\mu j\nu} \Delta_{i\mu j\nu}c^\dagger_{i\mu\uparrow}c^{\dagger}_{j\nu\downarrow}+{\rm h.c.},\\
H_{int}&=&U\sum_{i\mu\sigma}\langle
n_{i\mu\bar{\sigma}}\rangle n_{i\mu\sigma}+(U-2J_H)\sum_{i\mu\sigma} \langle n_{i\bar{\mu}\bar{\sigma}}\rangle n_{i \mu {\sigma}} \nonumber\\
&+&(U-3J_H)\sum_{i\mu\sigma} \langle
n_{i\mu\sigma}\rangle n_{i\bar{\mu}\sigma},
\end{eqnarray}
where the singlet pairing parameter is defined as $\Delta_{i\mu j\nu}=\frac{V}{2}\langle
c_{i\mu\uparrow}c_{j\nu\downarrow}-c_{i\mu\downarrow}c_{j\nu\uparrow}\rangle$ with $V=2$. In our model pairings only exist among the same orbitals and only along nnn links, leading to the S$_\pm$ paring order in the absence of $B$.
In this study we select $U=3.4$ and $J_h=1.3$ to systematically investigate the inhomogeneous states of Fe-based high-Tc.
With these parameters, theoretical results on 122 systems are qualitatively consistent with experimental measurements~\cite{s1,s11,s2,s3,s4,s5,s6,exkt,exy,exj,exk,exjx}, and the properties of the system are stable within a large range of $U, J_h$~\cite{s11}.
We note that the Hubbard $U$ of significant strength, as well as the exchange interaction, also plays a crucial role in the formation of the Cooper pair and charge modulations~\cite{a48,a49} in the models appropriate for copper based high-$T_C$ superconductors~\cite{a50}.

The total Hamiltonian $H_{tot}=H_t+H_{int}+H_{sc}$ can be solved via the following Bogoliubov-de Gennes (BdG) equations~\cite{f3,f6,s11,yangkun}
\begin{eqnarray}
\sum_{j\nu} \left( \begin{array}{cc}
 \mathbb{H}_{i\mu j\nu\sigma} & \Delta_{i\mu j\nu}  \\
 \Delta^{*}_{i\mu j\nu} & -\mathbb{H^{*}}_{i\mu j\nu\bar{\sigma}}
\end{array}
\right) \left( \begin{array}{c} u^{n}_{j\nu\sigma}\\v^{n}_{j\nu\bar{\sigma}}
\end{array}
\right) =E_n \left( \begin{array}{c} u^{n}_{i\mu\sigma}\\v^{n}_{i\mu\bar{\sigma}}
\end{array}
\right).
\end{eqnarray}\label{eq:ham}
where $\mathbb{H}_{i\mu j\nu\sigma}$ contains the matrix elements of $H_t+H_{int}$. The mean-field order parameters are obtained self-consistently by
\begin{eqnarray}
\Delta_{i\mu j\nu}&=&\frac{V}{2}\sum_n \left \{ u^{n}_{i\mu\uparrow}v^{n*}_{j\nu\downarrow}[1-f(E_n)]-u^{n}_{j\nu\uparrow}v^{n*}_{i\mu\downarrow}f(E_n) \right \},\\
M_i&=&\frac{1}{2}\sum_{\mu} \left (\langle n_{i\mu\uparrow}\rangle-\langle
n_{i\mu\downarrow}\rangle \right ),
\end{eqnarray}
in which
\begin{equation}
\label{eq:occ}
n_{i\mu} =\sum_{n} \{|u^{n}_{i\mu\uparrow}|^{2}f(E_n)+|v^{n}_{i\mu\downarrow}|^{2}[1-f(E_n)]\}.
\end{equation}
Here, $f(E_n)$ is the Fermi distribution function.
The kinetic energy of the system can be expressed by the sum $E^{kin}=\Sigma_{\sigma i\mu j\nu}(\xi^{\sigma}_{i\mu j\nu}+\xi^{\sigma*}_{ i\mu j\nu})$, where
$\xi^{\sigma}_ {i\mu j\nu}=-t_{i\mu,j\nu}\langle c^{\dag}_{i\mu \sigma}c_{j\nu \sigma} \rangle$.
To take into account the directional differences, we also define $\Delta_i \equiv \frac{1}{8}\sum_{\tau \mu}\Delta_{i\mu,i+\tau\mu}$, $\xi^{\sigma}_{i d}\equiv\frac{1}{16}\sum_{\tau \mu \nu}\xi^{\sigma} _{i\mu,i+\tau\nu }$ with $\tau=\pm\hat{x}\pm\hat{y}$, and $\xi^{\sigma}_{i xy}\equiv\frac{1}{8}\sum_{\tau \mu} \xi^{\sigma}_{i\mu,i+\tau\mu}$ with $\tau=\pm \hat{x},\pm \hat{y} $.

We perform numerical calculations on a $N_x\times N_y=24\times 48$ lattice with periodic boundary conditions, which yields
$24\times 48\times2\times2 = 4608$ eigen-energies $E_n$ with band and spin indices taken into account.
We choose $M_x\times M_y=20\times 20$ supercells to calculate the LDOS $\rho_i(\omega)=\rho_{i\uparrow}(\omega)+\rho_{i\downarrow}(\omega)=\frac{1}{M_x M_y}\sum_{n \mu {\bf k}}{|u^n_{i \mu \uparrow{\bf k}}|^2\delta(E_{n,{\bf k}}-\omega)+|v^n_{i \mu \downarrow{\bf k}}|^2\delta(E_{n,\bf k}+\omega)}$ for each spin species and their sum,
where the wave vector is taken as ${\bf k}_{\alpha}=-\frac{\pi}{N_{\alpha}}+\frac{2\pi i}{N_{\alpha} M_{\alpha}},\alpha=x,y$ in the supercell techniques with $E_{n,{\bf k}}$ being eigen-energies for a fixed wave vector.
In the calculation we regulate $\delta(x)$ by $\frac{\Gamma}{\pi(\Gamma^2+x^2)}$ with $\Gamma= 0.01$.

\begin{figure}
\centering
      \includegraphics[width=4.1cm]{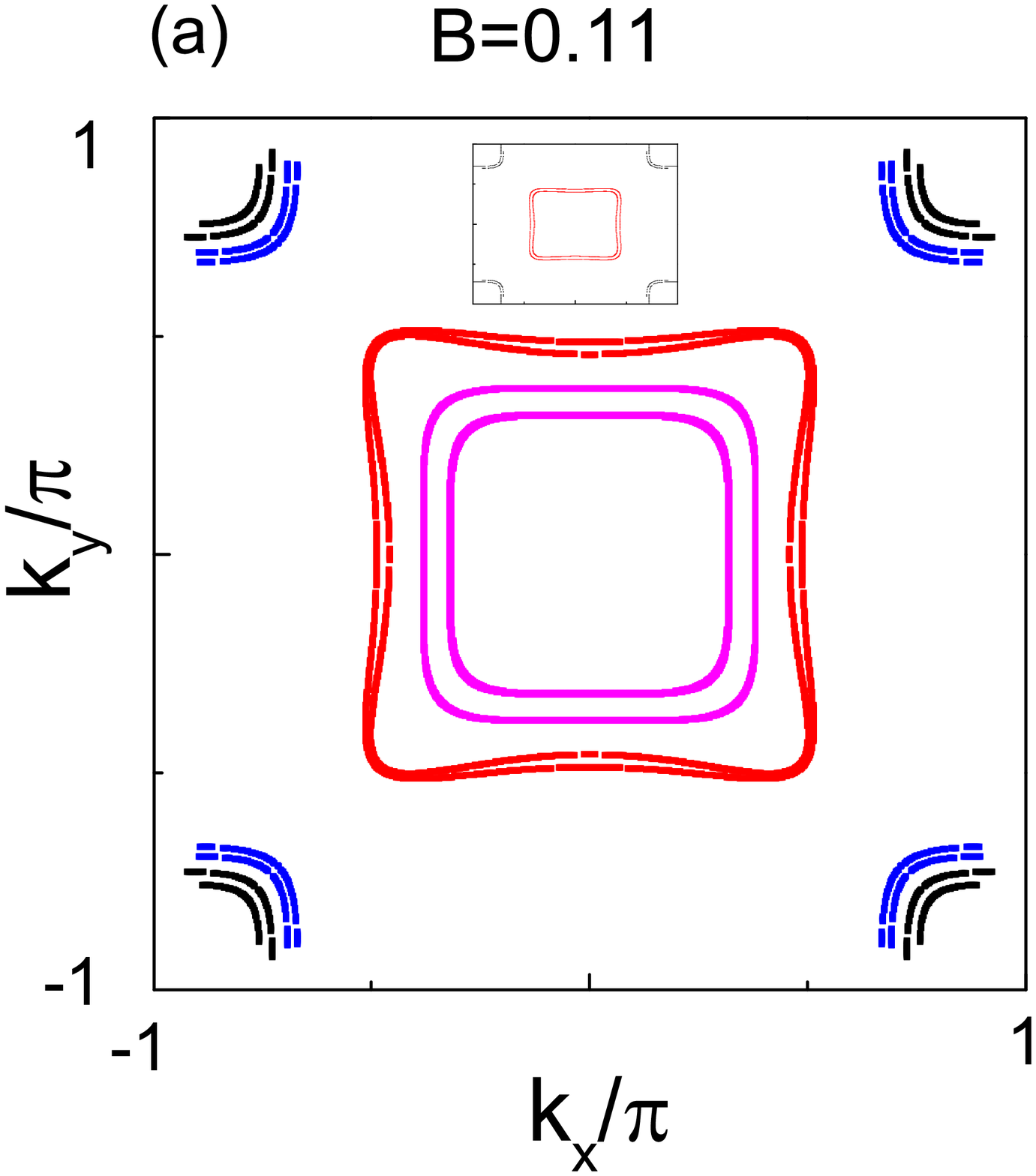}
      \includegraphics[width=4.3cm]{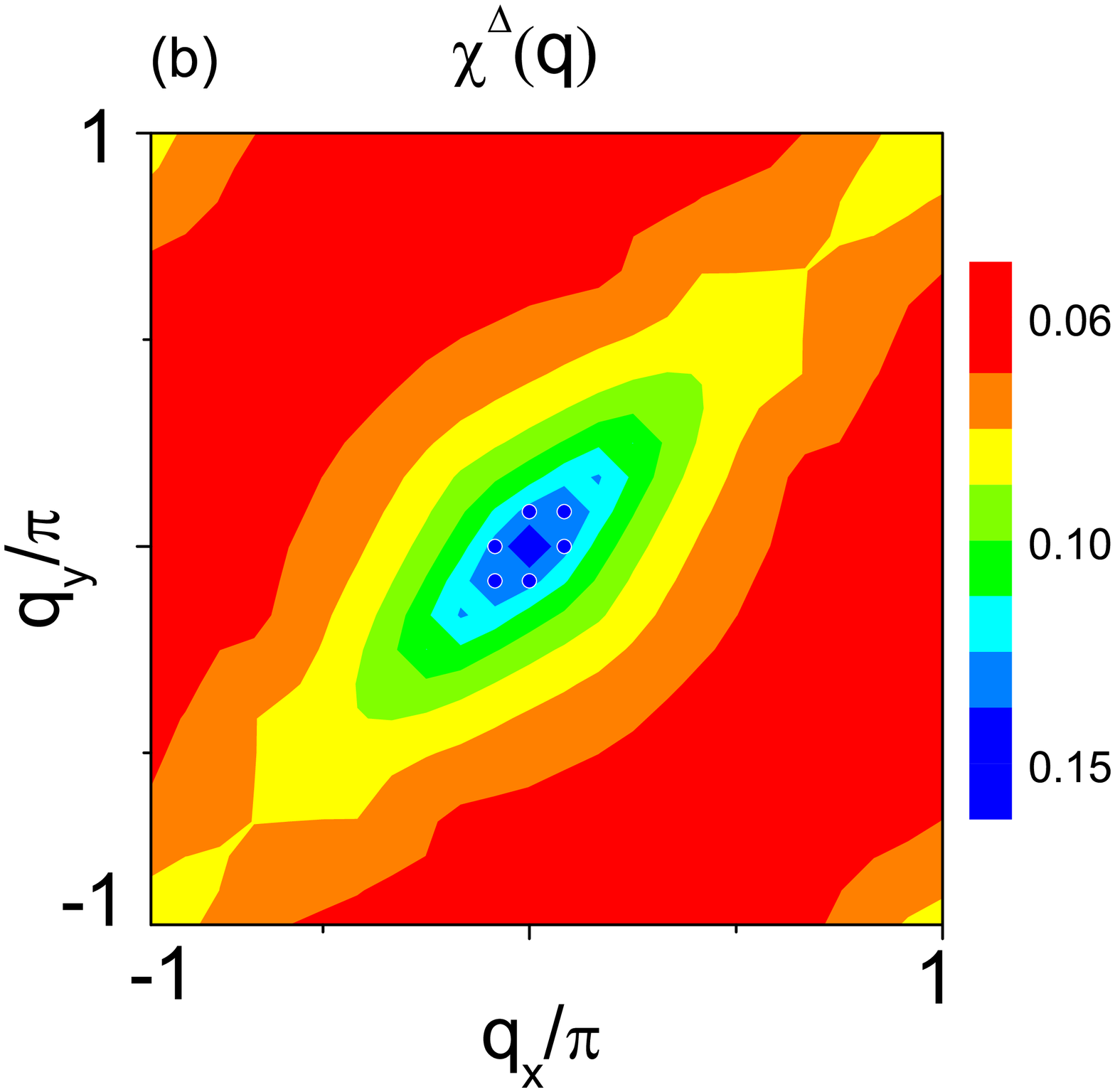}
\caption{(color online) (a) The Fermi surface topology at $B=0.11$ in the absence of paring. (b) Pairing susceptibility in the momentum space in the absence of $B$. The six high-susceptibility points are highlighted by blue dots. The corresponding Fermi surface topology is shown in the inset of (a) for comparison.
}\label{fig1}
\end{figure}
Fe-based superconductors have been argued to exhibit all the properties allowing the formation of the FFLO state~\cite{andimp}. After suppressing the SC order, we plot the Fermi surface topology in Fig.~\ref{fig1}(a) for $B=0.11$. Compared to the $B=0$ case [Fig.~\ref{fig1}(a) inset], we find that the disconnected sheets of the Fermi surface split due to the lift of spin degeneracy. Around the $\Gamma$ point, the spin-up electrons form the two inner loops (magenta), and the spin-down ones form the two outer loops (red). Meanwhile, the spin-up (down) electrons form the two outer (inner) pockets at the corners of the first Brillouin zone. Electrons of different spins from different sheets may be paired with non-zero total momentum, leading to the formation of the FFLO and other exotic states.

The retarded Green's function of Cooper pairs is defined as $G_{ \Delta_i \Delta^{\dag}_j }(t)=-i\Theta(t)\langle [\Delta_i(t), \Delta^{\dag}_j(0)] \rangle $ and its real-frequency Fourier transform reads $\chi^{\Delta}_{ij}(\omega)=\int dt G_{ \Delta_i \Delta^{\dag}_j }(t)\exp(i\omega t)$ with $\Theta(t)$ being the step function. In order to investigate the propagation of the net momentum in the density of Cooper pairs, we calculate the static Cooper pairs susceptibility in the absence of Zeeman field $\chi^{\Delta}(\textbf{q})=\lim_{\omega\rightarrow 0} \frac{-1}{N}\sum_{ij}\exp[\mathrm{i}{\bf q}\cdot(\textbf{R}_i-\textbf{R}_j)]\chi^{\Delta}_{ij}(\omega)$ where $\textbf{R}_i$ is the coordinate of site $i$ and $\textbf{q}$ is the momentum.
By using the equation of motion of the imaginary time correlation function in the Matsubara formalism, we obtain
\begin{eqnarray}\label{corr}
\chi^{\Delta}(\textbf{q})
            =\sum_{ m_1m_2,\omega=0}   \frac{
            \xi_{m_1m_2}(\textbf{q})\xi_{m_2m_1}(\textbf{q})[f(E_{m_1})-f(E_{m_2})]
            }{\mathrm{i}\omega+(E_{m_1}-E_{m_2})}.
\end{eqnarray}
The coefficient $\xi_{m_1m_2}(\textbf{q})=\sum_iD_{i,m_1 m_2}\exp(\mathrm{i}\textbf{q} \cdot \textbf{R}_i)$, where $D_{i,m_1 m_2}$ is given by the pairing order in the quasiparticle basis  $\Delta_i=\sum_{m_1,m_2}\gamma_{m_1}\gamma^{\dag}_{m_2}D_{i,m_1 m_2}$.
Fig.~\ref{fig1}(b) plots $\chi^{\Delta}(q_x,q_y)$ in the momentum space.
In addition to $\chi^{\Delta}(0,0)=0.15$, one finds more high-susceptibility points, e.g., with $\chi^{\Delta}(\pm0.262,\pm0.262) \approx \chi^{\Delta}(\pm0.262,0) \approx 0.14$, indicating that the system tends to form an FFLO state with a nonzero net momentum of Cooper pairs.

\section{Order parameters and Phase diagram }\label{results}

\begin{figure}
\centering
      \includegraphics[width=8.0cm]{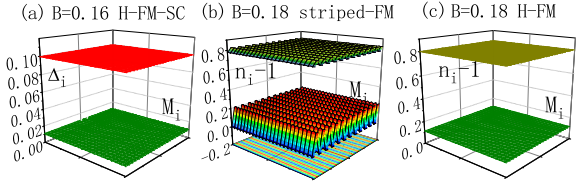}
\caption{(color online) (a) Spatial profiles of order parameters $M_i$ and $\Delta_i$ for the H-FM-SC state at $B = 0.16$.  Competing spatial profiles of $M_i$ and $n_i-1$ for (b) the striped-FM state and (c) the H-FM state are compared at $B = 0.18$. $M_i$ of the striped-FM state is also projected to the bottom plane for clarity. }\label{figorder}
\end{figure}\label{fig3d}

The model exhibits a variety of phases as the Zeeman field strength increases. If we only allow homogeneous solutions at zero temperature ($T=10^{-4}$ in numerical calculations), we find
the constant values of $\Delta_i=0.105$ and $M_i=0$ persist as $B$ increases, indicating a H-SC phase.  Beyond $B=0.13$, a homogeneous ferromagnetic superconducting (H-FM-SC) phase emerges where $\Delta_i$ has relatively high values in contrast to a small positive $M_i$. As shown in Fig.~\ref{figorder}(a), the SC order $\Delta_i=0.099$ is about 10 times larger than that of the FM order $M_i=0.01$ for $B=0.16$. The SC order disappears at $B\geq0.18$, and the system enters a trivial homogeneous ferromagnetic (H-FM) phase.

\begin{figure}
\centering
      \includegraphics[width=7.5cm]{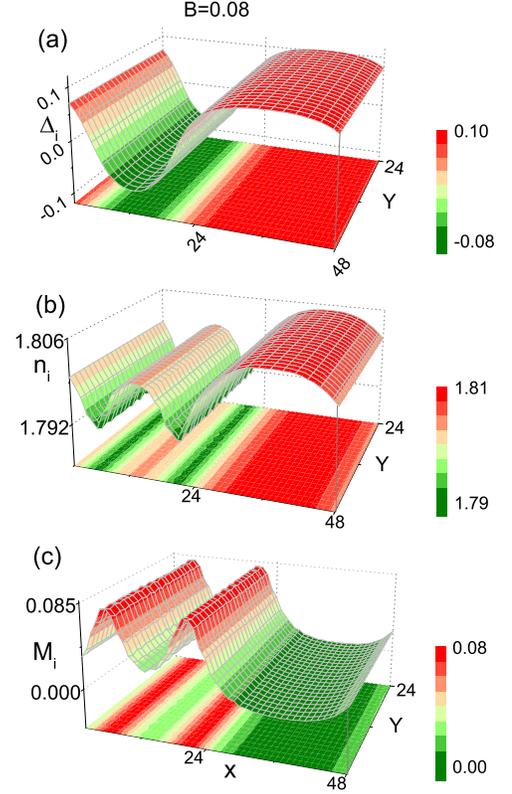}
\caption{(color online) Spatial profiles of the order parameters for the FFLO state at $B=0.08$. The contour plot of these order parameters are shown at the bottom. }\label{fig3}
\end{figure}

\begin{figure}
\centering
      \includegraphics[width=4.2cm]{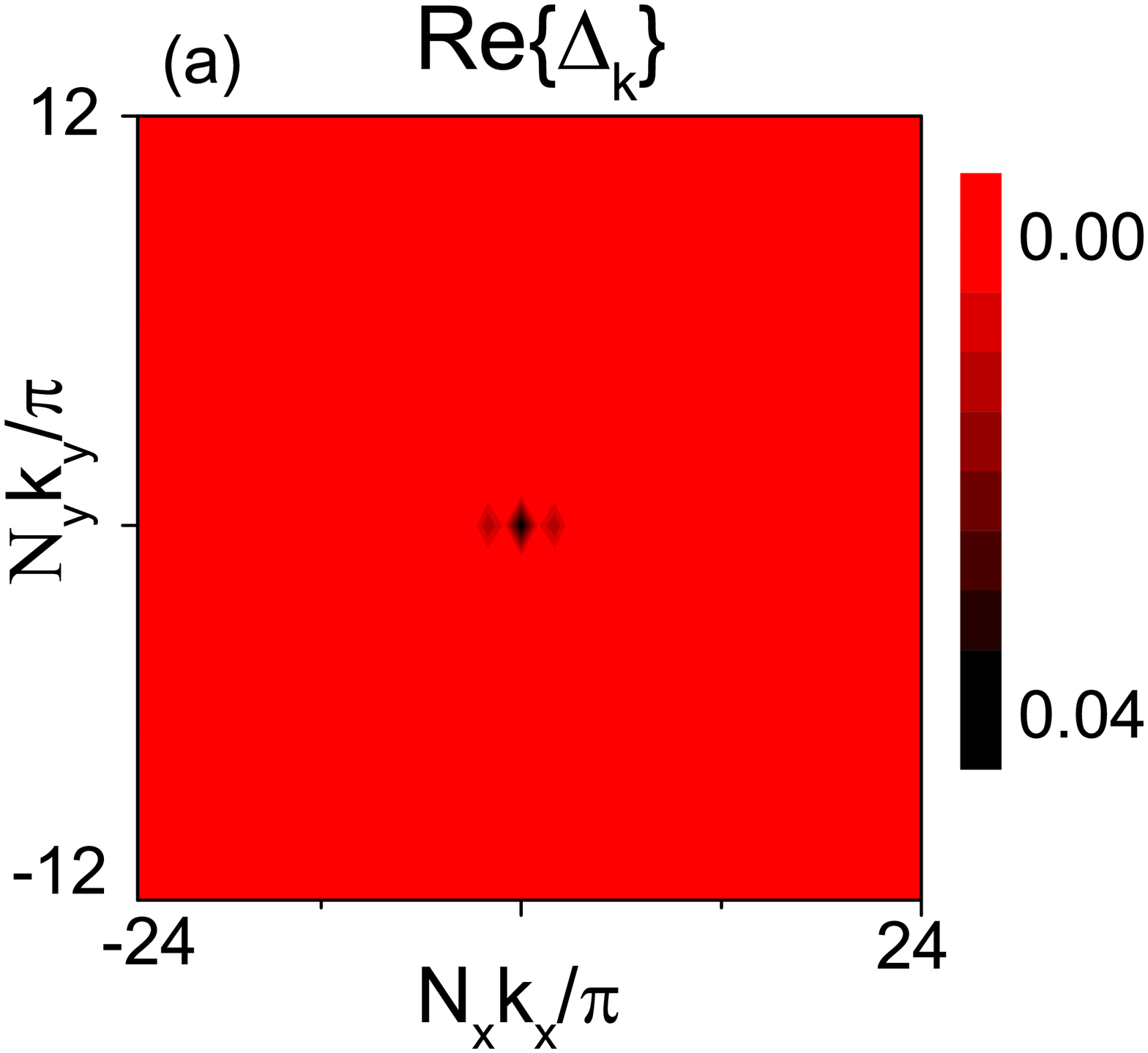}
      \includegraphics[width=4.2cm]{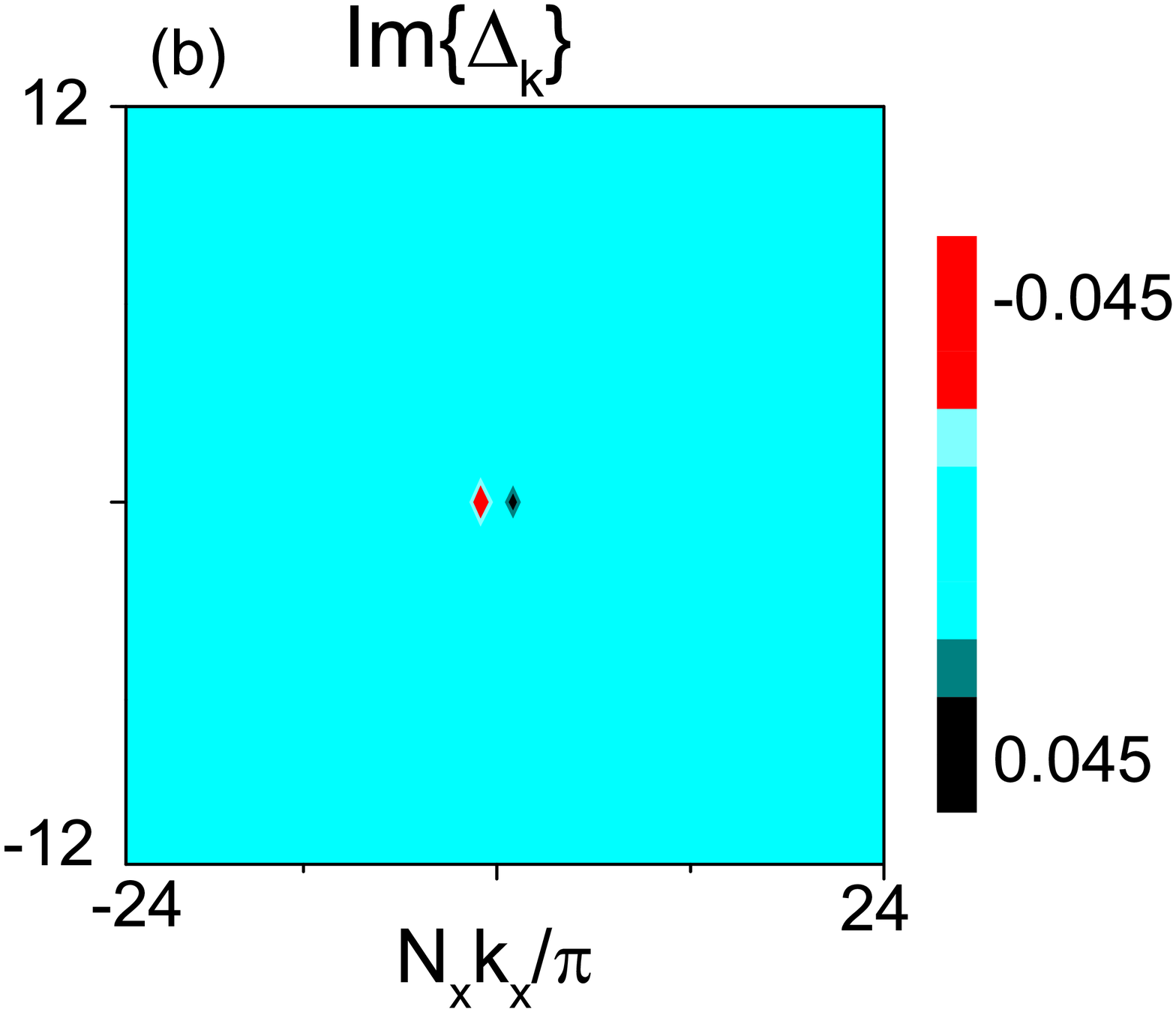}
\caption{(color online) (a) Real part and (b) imaginary part of the SC order $\Delta_\mathbf{k}$ in the momentum space at $B=0.08$.}\label{fig4}
\end{figure}

When we include inhomogeneous solutions, we find that the system can form an FFLO state, which is energetically more favorable than an H-SC state. Here, we restrict ourselves to the one-dimensional (1D) possibilities. In Fig.~\ref{fig3}(a), we show that the self-consistently determined $\Delta_i$ at $B = 0.08$ oscillates along $x$ direction with a nonzero mean, somewhat like a sine function. Figures~\ref{fig3}(b) and (c) show that the electron concentration $n_i$ oscillates in phase with $|\Delta_i|$, while $M_i$ competes with the SC order and oscillates out of phase with $|\Delta_i|$. In particular, $M_i$ maximizes when $\Delta_i=0$.
The sign of $M_i$ does not change, indicating an FM order.

To obtain the net momenta of the Cooper pairs we calculate the pairing order in the momentum space~\cite{yangkun} by $\Delta_k=\frac{1}{N_xN_y}\sum_i \Delta_ie^{-\mathrm{i}\mathbf{k}\cdot\mathbf{R}_i}$. The real part of the Fourier transform has a sharp peak ${\Delta_k}=0.039$ at $\mathbf{k}=0$, which is associated with the nonzero average of $\Delta_i$, and two weak peaks $\mathrm{Re} \{\Delta_k\}=0.015$ at $k_x=\pm0.26$, as shown in Fig.~\ref{fig4}(a). The imaginary part of $\Delta_k$, on the other hand, has two dominant components $\mathrm{Im} \{\Delta_k\}=\pm0.044$ at $k_x=\pm0.13$ as in Fig.~\ref{fig4}(b). By taking these dominant contributions, we can fit the SC order parameter by
$\Delta_i=0.039-0.088\sin(0.13x)+0.03\cos(0.26x)$ along the $x$ direction. 
If the Hamiltonian only contains the hopping term and the phenomenologically introduced SC pairing, the SC
order of the FFLO state is generally a single sinusoidal function
in the whole system. When an FFLO
state is caused by complex band structure and/or complex interaction,
the SC order can contain several harmonic terms as demonstrated in our study. Here, the oscillations of the sine and cosine terms nearly cancel each other for $x > 24$, leading to an apparent
suppression of the spatial variation of $\Delta_i$, as shown in Fig.~\ref{fig3}(a).

\begin{figure}
\centering
      \includegraphics[width=8.6cm]{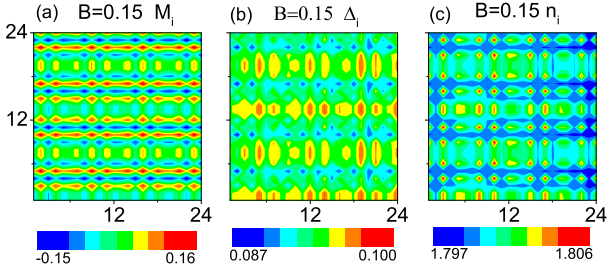}
\caption{(color online) Patterns of the order parameters (a) $M_i$, (b) $\Delta_i$, and (c) $n_i$ for the 2D TT-SC state at $B=0.15$.}\label{figtexture}
\end{figure}

At larger Zeeman field, the homogeneous H-FM-SC phase is replaced by a 2D TT-SC phase that features a complicated pattern of magnetic order with small average $\langle M_i\rangle$ but large $\langle |M_i| \rangle$, as shown in Fig.~\ref{figtexture}(a). For $B = 0.15$, we obtain $\langle|M_i|\rangle=0.0748$ and $\langle M_i\rangle=0.0074$. This antiferromagnetic-type of local spin arrangement leads to a revival of the SC order with a somewhat complementary pattern for $\Delta_i$ shown in Fig.~\ref{figtexture}(b). Although $\Delta_i$ varies spatially, the net momentum of the Cooper pairs is zero. Fig.~\ref{figtexture}(c) confirms that the electron concentration $n_i$ also has a texture pattern. We note that in both the FFLO and the 2D TT-SC states the amplitude fluctuation of $n_i$ is rather small (less than $0.015$).

\begin{figure}
\centering
      \includegraphics[width=8.8cm]{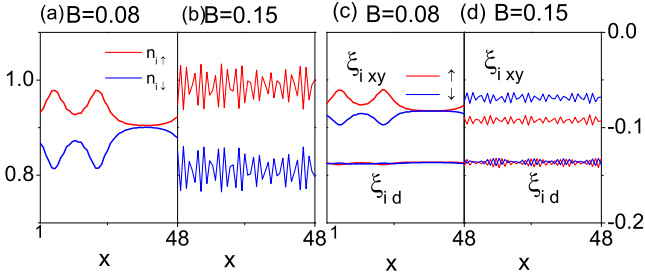}
\caption{(color online) Variations of spin-resolved $n_i$ along $y = 12$ for (a) the FFLO state at $B=0.08$ and (b) the 2D TT-SC state at $B=0.15$. For comparison, spin-resolved $\xi_{i xy}$ and $\xi_{i d}$ along the same cut are shown in (c) at $B=0.08$ and (d) at $B=0.15$.}
\label{xi}
\end{figure}

As electron concentration $n_{i\sigma}$ oscillates, the contribution to the kinetic energy varies as well. Fig.~\ref{xi} plots the variation of $n_{i\sigma}$ and $\xi^{\sigma}$ along $y=12$ for the FFLO and the 2D TT-SC states. In both cases $\xi^{\uparrow}$ and $\xi^{\downarrow} $ are out of phase, as shown in Figs.~\ref{xi}(c) and (d). We obtain the kinetic energy per site to be $E^{kin}=-2.842$ for the FFLO state and $E^{kin}=-2.834$ for the TT-SC state. The latter is almost identical to that of the H-SC state, while the energetically less favorable H-FM-SC state has an intermediate $E^{kin}= -2.838$. Therefore, the competition among the FFLO, TT-SC, and H-FM-SC phases manifests an intricate interplay among the kinetic energy and the magnetic and SC interactions.

When the Zeeman field is further increased, a striped-FM state, which has more favorable energy than the corresponding H-FM state, appears. As shown in Fig.~\ref{figorder}(b) for $B = 0.18$, $M_i$ and $n_i$ oscillate in phase, unlike in the FFLO and the 2D TT-SC phases. The striped-FM state retains the $2\times 1$ pattern as in the parent compound with its magnetic order $M_i$ varies in the range $[-0.01,0.22]$. The order parameters for the H-FM state at $B = 0.18$ is plotted in Fig.~\ref{figorder}(c) for comparison.
We point out that the striped-FM state cannot survive at $B>0.21$, where the trivial H-FM states are recovered.

\begin{figure}
\centering
      \includegraphics[width=8.8cm]{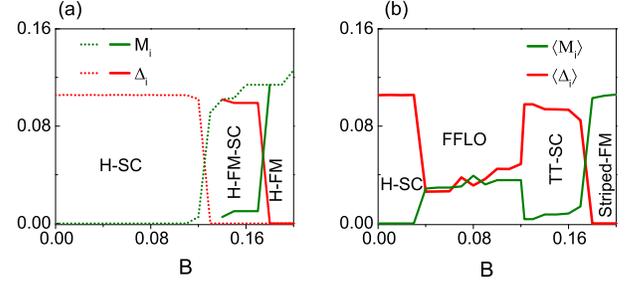}
\caption{(color online)
(a) Evolution of the order parameters $M_i$ and $\Delta_i$ for homogeneous solutions at $T=0$ for increasing $B$. The dotted lines are for the H-SC states and the H-FM state, while the solid lines for the H-FM-SC state.
(b) The averaged order parameter evolution for the energetically more favorable states, which include the inhomogeneous FFLO, 2D TT-SC, and striped-FM states.
}\label{figphase}
\end{figure}

We summarize the zero-$T$ phase diagram by plotting the average SC and magnetic order parameters in Fig.~\ref{figphase}. Panel (a) is only for the homogeneous states, while
(b) includes energetically more favorable inhomogeneous states. If we only allow homogeneous solutions, the SC order drops abruptly from $\Delta_i=0.105$ to zero at $B=0.13$, at which $M_i$ jumps from zero to a finite positive value, and the system goes into an H-FM state, as indicated by the dashed lines. But homogeneous $\Delta_i$ can revive at $B = 0.14$ at the expense of a much suppressed $M_i$, giving rise to  an H-FM-SC phase. At $B \geq 0.18$ the SC order fails to survive, and the H-FM phase rules with $\langle M_i \rangle=0.114$ for $B=0.18$.
The phase diagram including inhomogeneous solutions is shown in Fig.~\ref{figphase}(b). The FFLO phase is energetically favorable in the range $0.04\leq B \leq 0.12$. For example, at $B=0.04$ the inhomogeneous state has an energy $E \approx -2.76$, narrowly beating $E \approx -2.74$ for the H-SC state.
The 2D TT-SC phase with revived SC order, which beats the H-FM-SC phase in energy, appears in the range $0.12< B<0.18$. At $B=0.16$, for example, the 2D TT-SC state has $E \approx -2.80$ as oppose to $E \approx -2.78$ for the H-FM-SC state.
The SC order vanishes at $B = 0.18$, beyond which the system enters, first, the striped-FM phase and, finally, the H-FM phase.

\section{LDOS and energy distribution function}\label{dos}
\begin{figure}
\centering
      \includegraphics[width=8.6cm]{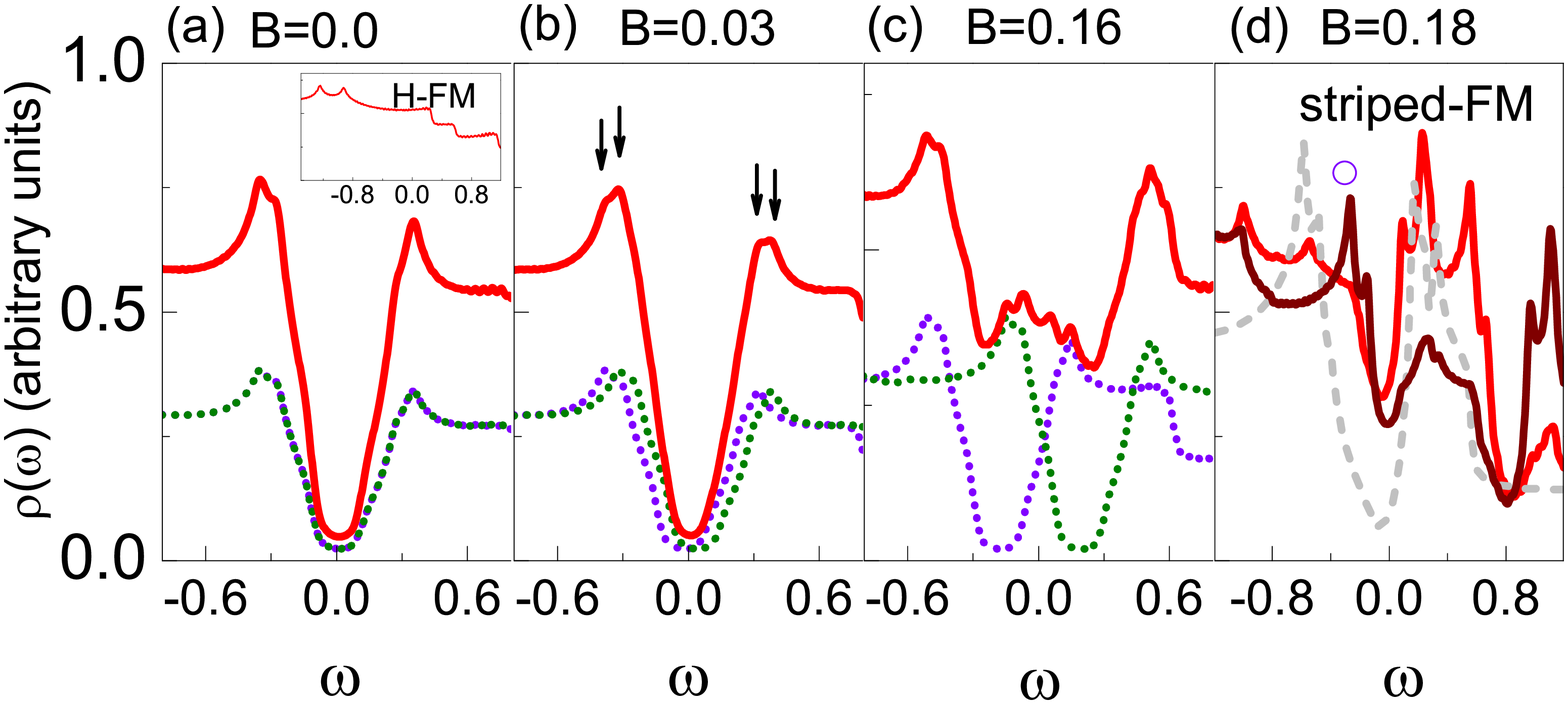}
\caption{(color online) The total (red), up-spin (purple), and down-spin (green) LDOS for
(a) the H-SC state at $B=0$, (b) the H-SC state at $B=0.03$, and (c) the H-FM-SC state at $B = 0.16$.
The inset of (a) shows $\rho$ of the H-FM state at $B=0.18$.
The black arrows in (b) indicate the locations of the double coherent peaks.
(d) The total LDOS for the inhomogeneous striped-FM state at $B = 0.18$ is shown
for sites of the maximum $M_i=0.22$ (red solid line) and for sites of the minimum $M_i=-0.01$ (brown solid line). The purple empty circle indicated the location of the SDW coherence peak.
For comparison, the grey dashed line shows the total LDOS of the parent compound.
}\label{1dos}
\end{figure}

\begin{figure}
\centering
      \includegraphics[width=8.6cm]{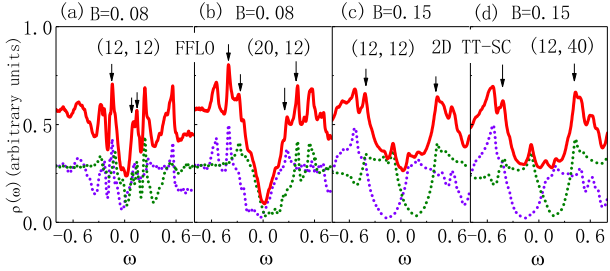}
\caption{(color online) LDOS of the inhomogeneous SC states (a) at site $(12,12)$ and (b) at site $(20,12)$ for the FFLO state at $B = 0.08$, as well as (c) at site $(12,12)$ and (d) at site $(40, 12)$ for the 2D TT-SC state at $B = 0.15$.
The black arrows indicate the locations of the coherent peaks.
Note that the two left peaks merge into one in (a).
}\label{ffdos}
\end{figure}

Figure~\ref{1dos} plots the LDOS $\rho$, as well as the spin-resolved contributions $\rho_{\sigma}$, in homogeneous phases and in the striped-FM phase.
The up- and down-spin LDOS are degenerate in the absence of Zeeman field, as shown in Fig.~\ref{1dos}(a).
In the H-SC state at $B = 0.03$ the U-shaped $\rho_{\uparrow}$ and $\rho_{\downarrow}$ are shifted by $B$ along opposite directions. Their sum $\rho$ remains to be U-shaped, as shown in Fig.~\ref{1dos}(b). As a result, two SC coherence peaks, indicated by black arrows, appear on either sides of the gap.
The LDOS for the H-FM-SC state, on the other hand, have in-gap resonance peaks due to the large shift of the two spin species $\rho_{\sigma}$ shown in Fig.~\ref{1dos}(c) at $B = 0.16$, even though each spin-resolved $\rho_{\sigma}$ retains its U shape.

In contrast, the SC coherent peaks vanishes in the striped-FM state, in which an SDW gap opens and asymmetric SDW coherence peaks appear. The picture is similar to that of the undoped antiferromagnetic parent compound at $B=0$, depicted by the grey dashed line in Fig.~\ref{1dos}(d).
We note that $\rho$ of the striped-FM state differs significantly from that of the H-FM state in the inset of Fig.~\ref{1dos}(a), which exhibits no coherence peaks.

In inhomogeneous SC states, the LDOS differs from site to site.
For illustration, we plot $\rho$ on two different sites for the FFLO state and the 2D TT-SC states.
We find that the spin-resolved $\rho_{\uparrow}$ and $\rho_{\downarrow}$ are no longer U-shaped in either phase.
The FFLO state features a V-shaped total $\rho$ as shown in Fig.~\ref{ffdos}(a) and Fig.~\ref{ffdos}(b) at $B=0.08$.
Still, there are two coherence SC peaks on either side of the Fermi energy; specially, the two low-energy peaks merge into one in Fig.~\ref{ffdos}(a).
On the other hand, the 2D TT-SC state at $B = 0.15$ features in-gap resonance peaks due to the large shift between $\rho_{\uparrow }$ and $\rho_{\downarrow}$, as illustrated in Fig.~\ref{ffdos}(c) and Fig.~\ref{ffdos}(d).  Our investigation show that europium-based iron pnictides can, possibly, have several homogeneous and inhomogeneous FM-SC states, thus the scanning tunneling microscopy (STM) measurements can render quite different results, which need to be understood by both the total LDOS and the spin-resolved LDOS.

\begin{figure}
\centering
      \includegraphics[width=8.8cm]{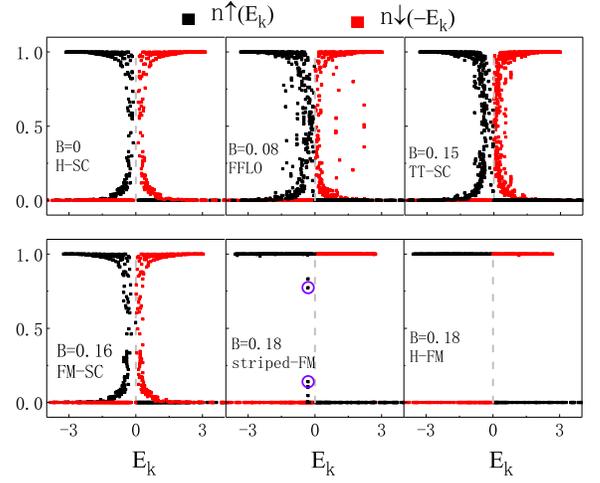}
\caption{(color online) Electron occupation $n_{\sigma}(E_k)$ for various phases, as the SC and FM orders compete with each other with increasing $B$.
}\label{plotuv}
\end{figure}

To further understand the SC phases, we plot the energy distribution function $n_{\sigma}(E_k)$ for various phases in Fig.~\ref{plotuv}. Here, each point represents a single state. For the H-SC phase in the absence of $B$, the existence of partially occupied points demonstrates the formation of Cooper pairs by electron near the Fermi surface. The trend of $n_{\sigma}(E_k)$ resembles the result in the standard BCS theory.
The result appears to be similar for the H-FM-SC phase at $B = 0.16$, which is not the most energetically favorable state as we discussed in Fig.~\ref{figphase}.
However, we point out that the Fermi energy has different distances to the up- and down-spin branches, and the total LDOS behaves very differently, as the spin-resolved LDOS shift far away from each other.

As $B$ increases above 0.04, the distribution of the partially occupied levels scatters well beyond the vicinity of the Fermi energy, as a result of the intricate interplay of the magnetic and SC energies of the interacting system in the FFLO phase.
As $B$ further increases, more states appear in the vicinity of the Fermi energy, while the energy spread of the electron participating in the pairing decreases, as shown for the TT-SC state at $B = 0.15$.
To some extent, we can think that the increasing magnetic interaction has been balanced by electron-electron interaction, thus absorbed into the nontrivial change of the mean-field band structure.
The large spread of the distribution in the FFLO state suggests that the momentum spread of the pairing states is also large, implying a relatively short coherence length according to the uncertainty principle.
Therefore, the results indicate that the FFLO state is the most fragile SC state and will be destroyed first by thermal fluctuations.

When Zeeman energy is even stronger, the SC order is completely suppressed, and $n_{\sigma}(E_k)$ becomes a step function, as in the H-FM state at $B = 0.18$.
When inhomogeneous states are allowed, the SDW order emerges.
There are degenerate states with intermediate occupation at $E_k=-0.304$, which corresponds to the SDW coherence peak in Fig.~\ref{1dos} (d) indicated by the purple circle.

\begin{figure}
\centering
      \includegraphics[width=8.8cm]{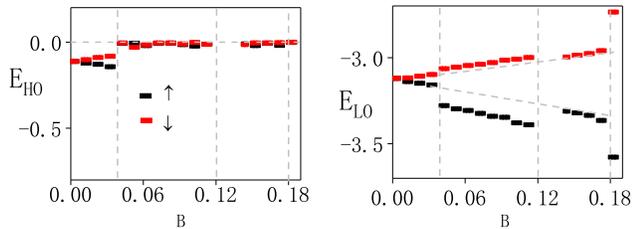}
\caption{(color online) The evolution of the highest and lowest occupied energies $E_{\rm HO}$ and $E_{\rm LO}$ for the ground states in Fig.~\ref{figphase}(b). Up- and down-spin levels are denoted in black and red, respectively.
}\label{ocp}
\end{figure}

An alternative way to understand the rich SC phases is to investigate the highest-occupied ($E_{\rm HO}$) and the lowest-occupied ($E_{\rm LO}$) energy levels for each spin species, which can be extracted from Fig.~\ref{plotuv}.
We plot their evolution with $B$ in Fig.~\ref{ocp} for the stable states in Fig.~\ref{figphase}(b).
For the H-SC states, the separation between up- and down-spin $E_{\rm HO}$ is exactly $2B$ as expected in a paramagnetic SC phase, as is the same for $E_{\rm LO}$.
In the FFLO phase, the hard gap near the Fermi energy disappears while the magnetic order appears, as indicated in Fig.~\ref{figphase}.
As a result, the up-spin energy levels gets further pushed down by the internal magnetic field and the gain in kinetic energy, as reflected in $E_{\rm LO}$.
In the 2D TT-SC states, $E_{\rm HO}$ behave similar to that of the FFLO states, but the separation of up- and down-spin $E_{\rm LO}$ returns to approximately $2B$.

\section{summary}\label{final}

We explore the zero-temperature phase diagram of europium-based iron pnictides in the presence of a Zeeman field.
In our mean-field study the field drives a transition in Fe-planes from an H-SC phase to inhomogeneous states with competing FM and SC orders.
As the field increases, the FFLO phase first emerges, followed by the TT-SC phase, which features 2D patterns of order parameter oscillations on shorter length scales.
In both inhomogeneous phases, spin-up and spin-down electrons fill up to the Fermi energy, and the hard SC gap give way to V-shaped spin-resolved LDOS.
The distribution of electron occupation for the FFLO and 2D TT-SC states
shows that electrons deep beneath the Fermi energy participate in the formation of Cooper pairs with short coherence length, unlike in the conventional BCS theory.

In our calculations the system can have more than one mean-field solutions due to the complex interplay among the Zeeman field, the SDW band structure of the parent compound, and the SC order.
In particular, homogeneous but energetically less favorable H-SC and H-FM-SC states compete with the FFLO phase and TT-SC phase, respectively.
The Zeeman field in the Fe-planes, which can arise from the FM ordering of the neighboring Eu-planes, is the crucial ingredient in our theory.
Though there is no direct experimental evidence that the FM order is strictly homogeneous, our model should also work for weak inhomogeneity, as the effective Zeeman field in the Fe plane averages the contributions from many spins in the spatially separated Eu planes.
Such interaction between drastically different but both dynamically nontrivial 2D layers may hold the key to understand the complex phase diagram of europium-based iron pnictides observed in experiments.

\section{acknowledgements}
H.X.H. would like to thank Yi Gao and Tao Zhou for helpful discussions. This work was supported by the National Natural Science Foundation of China Grant No. 11774218 and the Strategic Priority Research Program of Chinese Academy of Sciences through Grant No. XDB28000000.

\end{document}